# Chinese Medical Device Market and The Investment Vector

Weifan Zhang[a], Rebecca Liu[b], Chris Chatwin[a*]

[a] *School of Engineering and Informatics, University of Sussex, Falmer, Brighton, BN1 9QT, UK*
[b] *Management School, Charles Carter Building, Lancaster University, Lancaster, LA1 4YX, UK*

**ABSTRACT**

China has attracted increasing amounts of foreign investment since it opened its doors to the world and whilst many analysts have focused on foreign investment in popular areas, little has been written about medical device investment. The purpose of this article is to analyze the status of China's medical device market from the perspective of the healthcare industry and its important market drivers; the study reveals that the medical device market has significant growth potential. This article aims to identify and assess the profitable sectors of medical device technologies as a guide for international companies and investors.

[*] Corresponding author. Tel.: +44 1273 678901 or +44 1273 678907.
  E-mail address: C.R.Chatwin@sussex.ac.uk (C. Chatwin), wz41@sussex.ac.uk (W. Zhang), r.liu1@lancaster.ac.uk (R. Liu)

1. **Introduction**

The economy in China has experienced 30-years of rapid growth and obtained remarkable success ever since the great reforms and opening-up policy. This reform in China represents a fundamental change and overhaul of China's: value system, law making, institutional infrastructure, and socio-economic structure; this has been axiomatic for mobilizing foreign investment. While economic cycles and uncertainties have affected several global industries in the last few years, the medical device industry has gained from the benefits of earlier investments and delivered incomparable improvements in the quality of people's lives in the developed countries (Panescu, 2009). There is an urgent need for China and the world to better understand the Chinese medical device market and how it compares with other industries in China. One powerful development supporting a positive hypothesis is that people are paying more attention to their health due to the improved quality of life. Another is that treatment of some serious diseases requires high quality medical devices. So the research question arising is: What are the important features of the Chinese healthcare market indicating the investment potential of the Chinese medical device market?

Whilst there is some literature on the medical device industry and technology investment, it is narrowly focused and more general investment data is scarce. It is on the measure of profitability that the medical device industry truly stands out, as it has a more consistent rate of growth than nearly every other industry (Kruger, 2005). The medical device industry is increasingly significant in the contribution it makes to countries' economic productivity (Faulkner, 2009). If the market is being fully exploited, the principal difference between any two production locations is likely to be labor costs. When product standardization and market saturation give rise to cost pressures and price competition in developed countries, those companies that want to increase their export business, transfer technology and investment abroad, which means the developed countries shift production to the developing countries (Hill, 2009). The developing countries technological capabilities are often improved and expanded, this results in a stable technical ability enhancement via continuous accumulation. Technological capability enhancement is strongly correlated with foreign investment growth and vice versa (Cantwell, 1999; Narula, 1996). When a technological industrial firm in a developed country invests in a developing country (this country should gradually assimilate the economic structure), not only is the knowledge content of the investment important but it will also be directed towards a growth industry (Dunning, 1972). This explains why many US, German and Japanese companies have been able to create an important presence in the newer Chinese medical devices industry; this is because these countries were much more advanced in this technology area than China. China has substantial size and a high growth rate in its consumer market, which drives sales of service from healthcare, this is the reason why China is attractive to foreign investors (Cui & Liu, 2000). As an emerging market, China offers long-term growth opportunities that no longer exists in relatively saturated and highly competitive developed markets (Sakarya, 2007).

Investment clearly implies strong consumption now, with the expectation of more consumption at a later time (Au & Au, 1983). Despite the low rates of return on investment, China has attracted a great deal of foreign investment in many industries, while other selected countries were not attracting investment, or are even losing foreign capital in Hsiao's research (Frank Hsiao & Hsiao;, 2004). With adopted foreign capital, management know-how and trained labor, China possesses the capacity to absorb high technology industry, especially in the medical device industry (Liu & K. Daly, 2011). China has been the fastest growing economy, expanding at 10.0 percent annually, driven by exports and investment. High priority is given to transform the economic structure from an export driven to a consumption driven economy during the period of the "National 12$^{th}$ Five-Year Plan".[1] (Frost & Sullivan, 2010)

Medical devices can guide optimized medical intervention plans, treat people's illnesses and reduce the discomfort caused by the disease. High quality, well designed medical devices provide safe and effective clinical care for patients (Martin, Norris, Murphy, & Crowe, 2008). Of all the new high technology industries, medical device investment has the most potential. The healthcare industry has emerged as a driver of economic growth supported by government initiatives and increased investment. The size of the medical device market continues to exhibit a rapid growth trend in China, the growth rate recently reaching 23%, the market scale has reached approximately 19 billion U.S. dollars (USD) (Medical Economic News, 2010). However, the global medical device market is already over USD 300 billion, it was about 16 times more than China's in 2010 (World Health Organization, 2011). The US, Japan, Germany, France and Italy account for 13.1% of global population and 76% of global medical device use, conversely, the five most populous countries: China, India, Brazil, Indonesia and Pakistan account for about half of the global population but only 4.4% of medical devices used in the world (Zapiain, 2011). According to the Frost & Sullivan forecast, the growth rate of global medical devices will increase 4% to 6% annually in the next few years and China's entire medical device market is expected to double, reaching USD 53.7 billion by 2015 (Frost & Sullivan, 2010). Moreover, our study also finds that the recent development plans of the Chinese government are aimed at expanding the healthcare infrastructure across the rural-urban space. Opportunities exist in pharmaceutical, medical devices and equipment, outsourcing and biotechnology.

Since China opened the door to the world, its economic reforms have attracted more and more foreign investment. During this economic period there has been massive research into the impacts of foreign direct investment (FDI). A literature review of FDI and trade in health services reveals FDI involves the increased commercialization of the health care sector (Smith, 2004). Researchers used the bibliometrics analysis of many literature reviews, to show the main research streams on FDI into China (Fetscherin, Voss, & Gugler, 2010). Although there is growing interest in investment in China, analysis of the medical device area has been slow to develop, but the sector is growing in importance with the improved realization that investment in the medical technology area, especially medical devices, produces exceptional value.

---

[1] Five-Year Plan (FYP) is a series of social and economic development initiatives, which renews every five years. The Five-Year Plan was shaped by the Communist Party of China, who plays a leading role in mapping strategies for China's economic development, setting growth targets and launching reforms. First FYP: 1953-1957, the rest can be done in the same manner. So 11$^{th}$ FYP is from 2006-2010 and 12$^{th}$ FYP is from 2011-2015.

Based on these facts, this article aims to explore the potential and the investment opportunities in the Chinese medical device market. The purpose of this article is twofold. Firstly, it illustrates the strength of the Chinese medical device market from the perspective of time lines through a comparative analysis of its past and present performance. Secondly, it seeks to understand the medical device market in China by conducting a comparative study of secondary data.

This article contributes on several fronts as literature on investment in medical devices is scarce; this article contributes to the literature on both investment and the medical device market. It extracts technology investment theories from foreign investment theories. The medical device market has become a promising market in the world in recent years, especially since the Chinese government is more concerned about healthcare than before. Relevant policies on medical device investment will be helpful to investors doing business in this area. More and more business research will focus on the medical device investment area in the future. Few papers focus on investment in the medical device market, with little attention focused on medical device regulation. To bridge the gap, this article sheds some light on more detailed categories of medical devices that have promising investment potential. Using neural network time series predication, we are able to provide some insights on the future investment opportunities in the Chinese medical device market.

The structure of this article is organized as follows: Section One briefly introduces China's current economic situation and literature on foreign investments in the medical device area. Section Two reviews the Chinese medical device market from an economic perspective. Section Three illustrates the data collection and methodology of this paper. Section Four provides a detailed description of the important drivers for Chinese medical device market investment. Section Five discusses the findings of the research and Section Six provides the conclusions of the paper.

## 2. Background to the analysis --- Chinese medical device market

With the rapid growth of its economy, China has expanded its healthcare expenditure. According to the World Bank data, China accounts for only 4% of global healthcare expenditure and its healthcare expenditure, as a percentage of gross domestic product (GDP)[2], accounted for just 5.1% of the economy in 2009 (The World Bank, 2011a). During the years from 2002 to 2009, China's total annual healthcare expenditure gradually increased. Figure 1 shows total healthcare expenditure as a percentage of GDP from 2002 to 2009. By comparison, in 2009, total healthcare expenditure, as a percentage of GDP for the: US accounts for 17.7%; Germany-11.7%; UK-9.7%; Japan-9.5%, even Brazil is 8.8%, which is higher than China (The World Bank, 2011a). The above data suggests that China has the capacity to substantially improve its healthcare service. A significant factor is that China has the largest population in the world (1,331 million in 2009), which is nearly 4 times that of the US (307 million) and about 21 times that of the UK (62 million) in 2009 (The World Bank, 2011b). More specifically, in 2009, total health expenditure per capita[3] for the US was USD 7,990 and UK was USD 3,445 while China was only USD 191 (The World Bank, 2012b). The medical fee-for-service system in China leads to many Chinese people needing to pay for the services from their own pocket rather than being paid by the government as in the UK and Russia. The medical fee-for-service system creates barriers to seeking adequate quantity and quality of care for poor families and individuals (Garcia-Diaz & Sosa-Rubi, 2011). Research reports that private spending to pay for the medical services has skyrocketed, but the percentage of health care costs paid for by the government has decreased in China (Schulze, 2001). The Chinese government is trying to make medical services for people affordable, especially for the low-income people. The government has determined to carry out healthcare reform, and invested 850 billion RMB (about USD 140 billion) in healthcare systems from 2009 to 2011 (Ministry of Finance of the People's Republic of China, 2009). According to the government report, the reform has five aspects. (1) establish a basic medical security system and medical insurance system; (2) establish a national drug system; (3) improve the basic medical and health service system; (4) improve the equality of basic public health services; and (5) promote the reform of public hospitals. All this will inevitably lead to spending on capital goods, most notably medical devices. (Wong, 2013). Despite these reforms, the Chinese government still needs to be vigilant both technically and politically due to the challenges of: affordability and accessibility to medical devices (Xu & van de Ven, 2013).

---

[2] Gross domestic product (GDP) is the market value of all officially recognized final goods and services produced within a country in a given period of time.
[3] Total health expenditure per capita is the sum of public and private health expenditures as a ratio of total population.

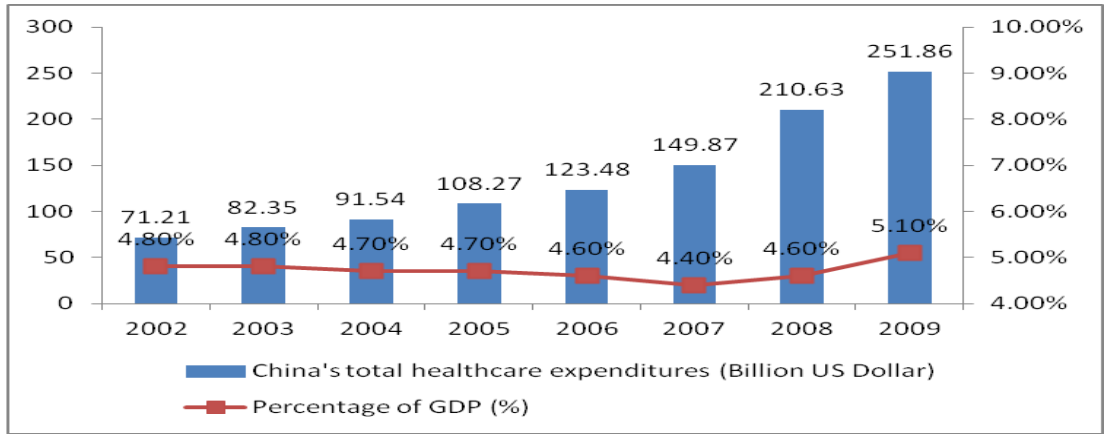

**Fig. 1.** China's total healthcare expenditures and percentage of GDP (2002-2009). Source: Chinese Health Statistics Digest 2010, (Ministry of Health of P.R.China, 2011)

The global medical device market is highly centralized (APCO Worldwide, 2010). The market share of the developed countries accounted for more than 80% of the global medical device market share (US: 42.4%, Europe: 33%, Japan: 11%) in 2011 (Espicom Business Intelligence, 2012). With superior know-how in technology and/or management, international companies are typically larger than domestic companies and have a competitive advantage due to the economies of scale (Hymer, 1976). According to Charles Hill and Vernon's product life cycle theory, the developed countries will export their production and technology from their relatively saturated market to the developing countries due to the market pressures and other competition in their established markets (Hill, 2009; Vernon, 1966; Vernon & Wells, 1986). Despite China only accounting for 3% of the global medical device market share (Espicom Business Intelligence, 2012), this study shows that the developing countries' medical device markets are experiencing rapid growth, especially in China. Increasing medical expenditure, rising healthcare consumption and health awareness improvements are all possible factors in promoting the development of the Chinese medical device market. The Chinese government's healthcare reform has injected additional "power" into the development of the medical device market. In fact, by the end of 2009, the Chinese medical device industry output value was expected to exceed 90 billion RMB, about USD 15 billion, the total percentage of GDP is 0.27% according to the National Bureau of Statistics of China, 2010. By the end of 2011, China's medical device industry output value was 110.3 billion RMB, total percentage of GDP is 0.29%. Figure 2 shows China's medical device industry output value and its total percentage of GDP, its value continues to climb from 2005 to 2010.

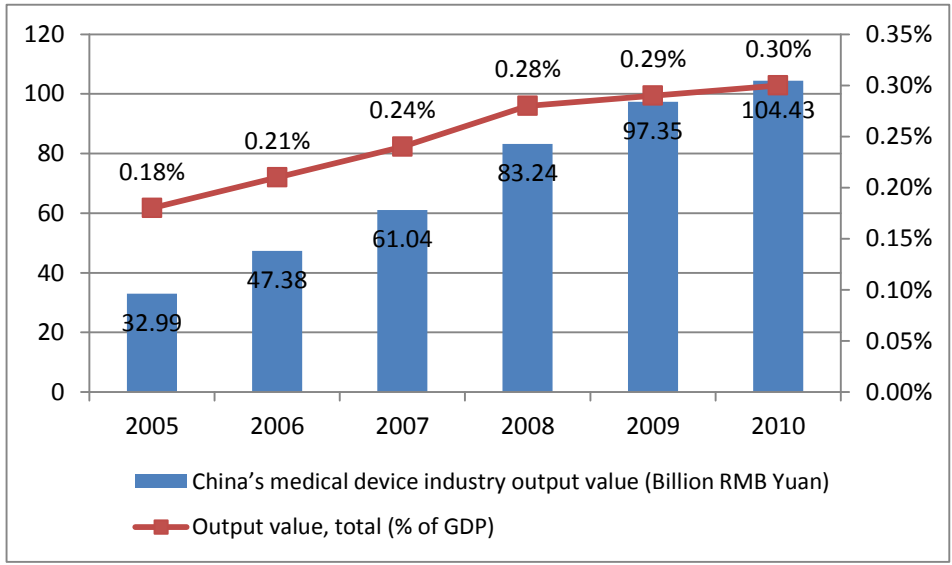

**Fig. 2.** China's medical device industrial output value and its total (% of GDP).
Source: National Bureau of Statistics of China, (National Bureau of Statistics of China, 2011a)

This study also suggests that the Chinese high-end medical device market is dominated by the US, Germany and Japan, it is dependent upon imports from these countries. The US is the most significant medical device market in the world. Close to 60% of all medical devices consumed around the world are produced by American companies (Kruger, 2005). According to the APCO Worldwide market analysis report, in China, 90% of value-added high-tech medical devices are foreign made, which accounts for 70% of China's medical device market (APCO Worldwide, 2010). These high-tech medical devices include CT, MRI, Ultrasound, X-ray, Implants and Assistive devices. Table 1 shows China's medical market trade statistics according to the China Chamber of Commerce for Import & Export of Medicines & Health Products (CCCMHPIE) in 2010. The overall trend of the Chinese healthcare market shows that export value is higher than import value; hence the export value of the pharmaceutical and medical device industry is higher than the import value. However, the import value of medical diagnosis and treatment devices is higher than the export value, which took a 29.05% share of total import volume in China. By comparison, medical dressings, disposable products, health protection and recovery products, dental equipment and materials, total only 6.8% of import volume, which is only one-quarter of the import volume of the medical diagnosis and treatment sector. China has a greater demand for medical diagnosis and treatment devices. This is why China's medical device market is dominated by the US, Germany and Japan.

**Table 1**
China's import and export structure of medicines and health products, 2010.
(Unit: 10,000 USD)

| Trade name | Export Value ($*10^4$) | Export value growth rate annually (%) | Share in total export volume (%) | Import Value ($*10^4$) | Import value growth rate annually (%) | Share in total import volume (%) |
| --- | --- | --- | --- | --- | --- | --- |
| Total | 3,973,310.23 | 24.87 | 100 | 2,046,435.6 | 23.98 | 100 |
| 1.Traditional Chinese Medicine | 194,447.12 | 22.78 | 4.89 | 68,794.91 | 22.61 | 3.36 |
| 2.Pharmaceuticals | 2,393,002.43 | 28.17 | 60.23 | 1,244,083.7 | 20.53 | 60.79 |
| 3.Medical Devices | 1,385,860.67 | 19.83 | 34.88 | 733,556.93 | 30.45 | 35.85 |
| 3.1 Medical dressings | 468,750.68 | 11.95 | 11.8 | 20,776.8 | 25.63 | 1.02 |
| 3.2 Disposable products | 192,227.35 | 15.42 | 4.84 | 88,076.07 | 27.73 | 4.3 |
| 3.3 Medical diagnosis and treatment | 454,359.93 | 25.56 | 11.44 | 594,472.93 | 30.34 | 29.05 |
| 3.4 Health protection and recovery products | 241,640.74 | 30.87 | 6.08 | 14,937.21 | 83.83 | 0.73 |
| 3.5 Dental equipment and materials | 28,881.97 | 16.51 | 0.73 | 15,293.93 | 21.37 | 0.75 |

Source: CCCMHPIE, 2011 (China Chamber of Commerce for Import & Export of Medicines & Health Products, 2011a)

More specifically, Table 2 illustrates the trade statistics for medical devices in China in 2010. The total export value of medical devices reached USD 13.86 billion in 2010, while the total import value reached USD 7.3 billion. North America and Asia are the main export target areas for China, which accounted for 29.23% and 33.7% of the total export volume; the U.S. and Japan are the main export target countries, which absorbed 27.91% and 10.39% of the total export volume respectively. Europe and North America are the main exporters to China, which accounted for 39.01% and 31.41% of the total import volume. Germany and the U.S. are the main importing countries, which provide 17.34% and 30.71% of the total import volume.

Therefore, according to Table 1 and Table 2, China export medical products such as medical dressings, disposable products, health protection or recovery products and dental equipment or materials to North America and Asia while it imports medical diagnosis and treatment devices from North America and Europe.

**Table 2**

China's import and export markets of medical devices, 2010.     (Unit: 10,000 USD)

| Country | Export Value ($*10^4$) | Export value growth rate annually (%) | Share in total export volume (%) | Import Value ($*10^4$) | Import value growth rate annually (%) | Share in total import volume (%) |
|---|---|---|---|---|---|---|
| Total (All countries) | 1,385,860.67 | 19.83 | 100 | 733,556.93 | 30.45 | 100 |
| Asia | 466,995.82 | 12.87 | 33.7 | 193,877.25 | 29.28 | 26.43 |
| Europe | 363,106.46 | 18.53 | 26.2 | 286,137.63 | 34.38 | 39.01 |
| North America | 405,103.32 | 24.48 | 29.23 | 230,378.6 | 25.81 | 31.41 |
| 1.  U.S. | 386,760.61 | 24.72 | 27.91 | 225,263.7 | 26.48 | 30.71 |
| 2.  Germany | 77,806.39 | 14.72 | 5.61 | 127,173.94 | 35.43 | 17.34 |
| 3.  Japan | 144,007.49 | -13.5 | 10.39 | 111,390.48 | 26.99 | 15.18 |

Source: CCCMHPIE, 2011 (China Chamber of Commerce for Import & Export of Medicines & Health Products, 2011b)

Foreign investment took place because of the product and factor market imperfections (Hymer, 1976). Researchers proposed the relationship between market imperfections and sustainable opportunities, by recognizing and combining known supply and demand elements of the market (Cohen & Winn, 2007). For the Chinese medical device market, with growth from many sources of demand (unmet clinical needs, aging population, disease profiles, etc.), medical diagnosis and treatment devices still have a great growth potential in China. China now has a fee-for-service healthcare system financed largely by payments from patients, employers and health insurance companies (Chen, 2001). However, many patients, especially high income ones are willing to pay more for better treatment. This is particularly obvious for patients with cancer, heart disease and cerebrovascular disease that need to use high-tech medical diagnosis and treatment devices or high-grade drugs, which are not affordable for low income people. For example, the average fees for a CT whole body scan is nearly 2500 RMB (about USD 400) in China, this is not a small expenditure for low income people, they always choose the most economic ways to treat their diseases. However, China's health institutions especially the first class hospitals have a strong demand for high-end diagnostic devices due to the rising number of visits and inpatients, changed disease profiles etc.

## 3. Data and methods

### 3.1. Data

The data sources are mainly collected from government reports: (1) Ministry of Health of the People's Republic of China (now called National Health and Family Planning Commission of the People's Republic of China); (2) National Bureau of Statistics of the People's Republic of China; (3) United Nations (UN) and their organizations and specialized agencies such as the International Monetary Fund (IMF), World Bank and World Health Organization (WHO); (4) Some non-commercial agents such as China Association for Medical Devices Industry (CAMDI), China Association of Medical Equipment (CAME) and Chinese Hospital Association (CHA). Other data is collected from independent consulting companies such as Frost & Sullivan and Espicom Business Intelligence. The World Bank is an important source of finance and provides significant technical assistance to developing countries. The WHO focuses on international public health. The purpose of using these data sources is to understand the vitality of Chinese healthcare expenditure and the requirements of the Chinese population, which is helpful to identify the medical device market investment potential in China.

Data extracted from the National Bureau of Statistics of the People's Republic of China (China Statistical Yearbook) and Ministry of Health of the People's Republic of China (China Health Statistical Yearbook) from year 2000 to 2011, is presented in Table 3.

Table 3
Chinese healthcare related data.

| Time | Number of hospital visit (Billion) | China 65+ population (Billion) | China total healthcare expenditures (Billion RMB) | Hospital quantity (Units) | China medical devices revenues (Billion RMB) |
|---|---|---|---|---|---|
| 2000 | 1.286 | 0.089 | 458.663 | 16,318 | 14.5 |
| 2001 | 1.250 | 0.091 | 502.593 | 16,197 | 17.3 |
| 2002 | 1.243 | 0.093 | 579.003 | 17,844 | 20.7 |
| 2003 | 1.213 | 0.095 | 658.41 | 17,764 | 24.7 |
| 2004 | 1.305 | 0.097 | 759.029 | 18,393 | 29.5 |
| 2005 | 1.387 | 0.099 | 865.991 | 18,703 | 35.3 |
| 2006 | 1.471 | 0.101 | 984.334 | 19,246 | 43.4 |
| 2007 | 1.638 | 0.103 | 1157.397 | 19,852 | 53.5 |
| 2008 | 1.782 | 0.105 | 1453.54 | 19,712 | 65.9 |
| 2009 | 1.922 | 0.107 | 1754.192 | 20,291 | 81.2 |
| 2010 | 2.040 | 0.11 | 1998.039 | 20,918 | 120 |
| 2011 | 2.259 | 0.113 | 2434.591 | 21,979 | 147 |

The China Statistical Yearbook datasets contain data from every mainland Chinese province and city except: Hong Kong, Macao and Taiwan. It also includes the most important Chinese statistical data, which is an annual digest of economic and social development. China's total healthcare expenditures and hospital numbers are extracted from the China Statistical Yearbook, 2013. The major data sources are obtained from annual statistical reports and sample surveys. The number of hospital visits are available from the China Health Statistical Yearbook, 2012. This study does not include lower level health institutions in the hospital quantity, such as Grass-roots Health Care institutions [4], Specialized Public Health Institutions [5] and other Institutions. This is because, in China, medical diagnosis devices are always located in hospitals as other health institutions do not have the resources to purchase these expensive medical devices. Data on the numbers of Chinese aged 65 and above were collected from the UN Population Division. China's medical devices revenues were extracted from CAMDI.

*3.2. Methods*

This study analyses the Chinese medical device market using a combination of qualitative (official statistics and secondary analysis) and quantitative research methodologies (regression analysis). In this article many data sets are collected from official statistics, which are created by government departments or other organizations. Although there is some limitation to secondary analysis and official statistics, many researchers still use this method to analyze markets because it saves cost and time, gives access to high-quality data and provides the opportunity for longitudinal and cross-culture analysis, etc. (Bryman, 2012). Some forms of official statistics are very precise, such as births and deaths data (Bryman, 2012) and population census data (Bryman & Bell, 2011). Data analysis makes the developing trends more clearly understandable. Trend analysis can be used to predict the future value of the market. More and more researchers have started to focus on the past and its importance for understanding the present situation and for predicting the future more effectively via in-depth historical analysis.

In the population section, this article uses Neural Network Time Series prediction, to predict the total Chinese population and the number of 65 year olds and above from 2011 to 2020. Neural Networks are widely used in many areas but few researchers have used it for population prediction. This article analyses the data trends from the past to the present using longitudinal analysis, which shows the importance and changing trends for medical devices demand.

The most important drivers of the Chinese medical device market are: number of hospital visits; population above 65 years old; total healthcare expenditures and the number of medical institutions. Linear regression analysis is an effective method for studying the relationships and relative importance of these drivers. To evaluate the Chinese medical device market investment potential, we need to understand its revenues. China's medical device revenues is the dependent variable 'y', the independent variables 'x' are: number of visits ($x_1$), Chinese population above 65 years old ($x_2$), Chinese total healthcare expenditure ($x_3$) and the number of Chinese hospitals ($x_4$). Suppose the dependent variable y is combined with one independent variable $x_i$ and two parameters, $\beta_0$ and $\beta_1$:

$$y_i = \beta_0 + \beta_1 x_i, \quad i=1,2,\ldots,n. \tag{1}$$

---

[4] Grass-roots Health Care institution include community health centre and station, sub-district health centre, village clinic, outpatient department, and, clinic (infirmary)

[5] Specialized Public Health Institution include Chinese Center for Disease Control and Prevention (CDC), specialized disease prevention and treatment institution, health education centre, maternal and child health centre, emergency centre, centre for blood collection & supply, centre for health supervision and centre for family planning service.

Microsoft Excel is used for correlation analysis. We also used MATLAB for correlation analysis and obtained the same results. Using the official data in Table 3; after correlation, we conclude the relationship between number of hospital visits ($x_1$) and China's medical devices revenues (y) is:

$$y_1 = -127.35 + 116.05\, x_1 \tag{2}$$

The multiple R is 0.98, which means that medical devices revenues have a very positive correlation with the number of visits. In summary, the correlation equation means that every one million change in the number of visits will cause a positive change of 116.05 million in medical devices revenues.

The same methodology is used for correlation analysis of the other factors. After calculation, the relationship between China's 65+ population ($x_2$) and medical devices revenues (y) is:

$$y_2 = -470.54 + 5236.43\, x_2 \tag{3}$$

The multiple R is 0.94, which means there is a very positive correlation between China's 65+ population and medical devices revenues. In summary, the correlation equation means that every one million change in China's 65+ population will cause a positive change of 5236.43 million in medical devices revenues.

The relationship between China's total healthcare expenditure ($x_3$) and medical devices revenues (y) is:

$$y_3 = -19.53 + 0.07\, x_3 \tag{4}$$

The multiple R is 0.99, which means every one billion change in total healthcare expenditure will cause a positive change of 0.07 billion in medical devices revenues. The relationship between hospital quantity ($x_4$) and medical devices revenues (y) is:

$$y_4 = -364.46 + 0.02\, x_4 \tag{5}$$

The multiple R is 0.91, which means every one million change in hospital quantity will cause a positive change of 0.07 million in medical devices revenues.

## 4. Analysis

### 4.1. Number of hospital visits

From 2000 to 2011, the compound annual growth rate (CAGR) of hospital visits was 5.26%. According to the previous analysis, the correlation equation (2): $y_1 = -127.35 + 116.05 \, x_1$ means that every one million change in number of hospital visits will cause a positive change of 116.05 million in medical devices revenues. This result shows that the number of hospital visits is a vital contributor to the Chinese medical device market. One more thing we need to be aware of is that, grass-roots health care institutions account for more than 90% of the total number of medical and health institutions in China. However, the CAGR of grass-roots health care institution visits was lower than the hospital visits at about 1.16%. The main reason for this situation is the technology level of the medical devices in these health care institutions was lower than in hospitals. With the recent reform of China's healthcare system, this situation is changing. Thus, the speed of upgrading medical devices will cause substantial growth in the coming years.

### 4.2. 65+Population in China

According to the previous analysis, the correlation equation (3): $y_2 = -470.54 + 5236.43 \, x_2$, means for every one million change in China's 65+ population will cause a positive change of 5236.43 million in medical devices revenues. It is clear that population, especially the 65+ population, is the most important driver of the medical device market.

#### 4.2.1. Population and aging in China

The UN identifies, populations who have reached the age of 60 years as "older population" (Huber, 2005). Moreover, the UN considers a country to be aging when 10% of their total population is aged over 60 or 7% of their total population is aged over 65 (Zhang, 2012). According to the National Bureau of Statistics of China's Sixth National Population Census[6] in 2010 (National Bureau of Statistics of China, 2011b), the total population of China reached about 1.37 billion, where people aged 60 and above accounted for about 0.18 billion (13.26% of the total population), people aged 65 and above accounted for about 0.12 billion (8.87% of the total population). However, according to the IMF and UN statistics, China has 0.11 billion people that are over 65, which is 8.19% of the total population (1.34billion). Which means that China has entered the "aging society" category.

We can see the changing trend of the Chinese population between 1980 to 2010

---

[6] Peoples Republic of China have six national population census before. The first one was in the year of 1953, second one was in 1964, third one was in 1982, then 1990, 2000 and 2010.

from the IMF World Economic Outlook database, 2012 (International Monetary Fund, 2012), see Figure 3. In addition, it is necessary to understand China's profile in respect of the "aging society". There is no direct data on the percentage of 65 years olds and above from 1980 to 2010, but we can extrapolate how many 65 years olds and above there are from the UN Population Division (United Nations, 2011). After calculation, we can evaluate that the approximate percentage of 65 year olds and above in China from 1980 to 2010, see Figure 4. The IMF and UN data is given in Figure 3 and Figure 4. All the data is shown in Appendix B, Table B.

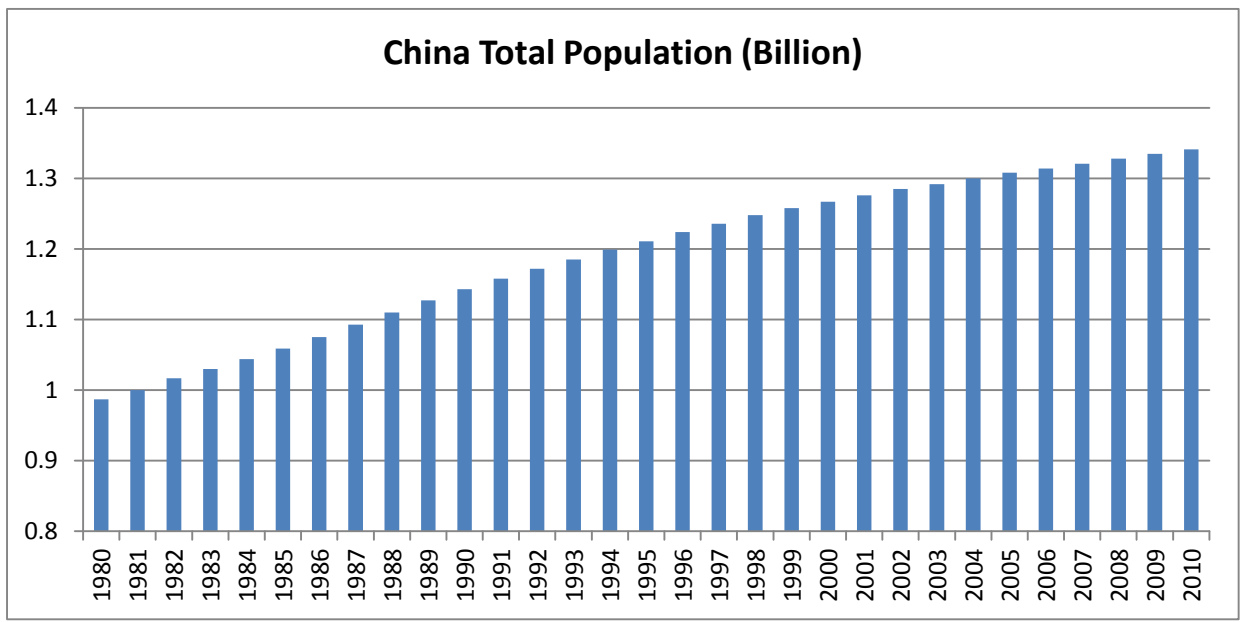

**Fig. 3.** Chinese population from 1980 to 2010.
Source: IMF, 2012 (International Monetary Fund, 2012)

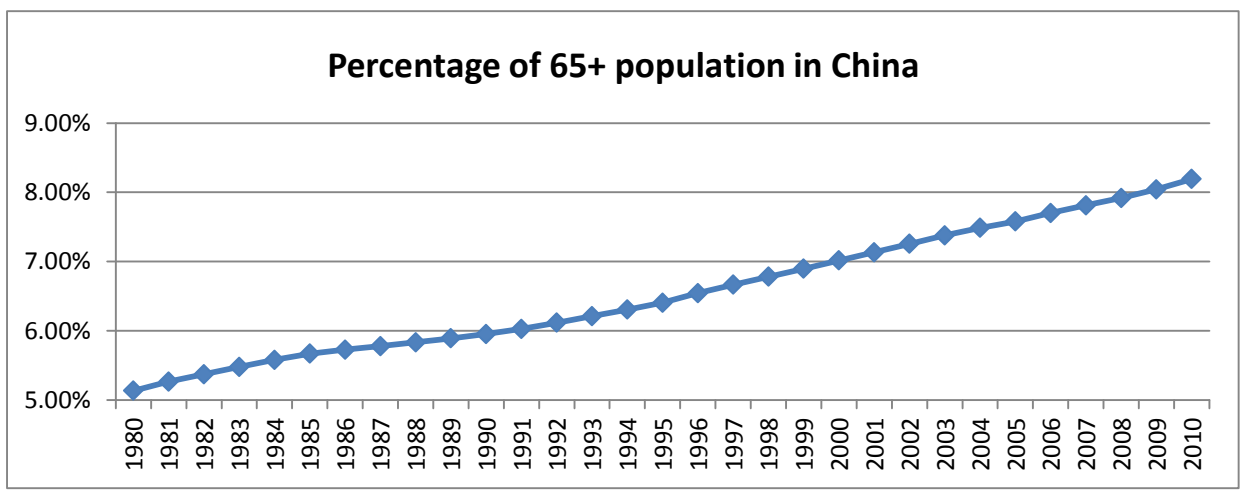

**Fig. 4.** Percentage of 65 year old and above population in China from 1980 to 2010.
Source: UN, 2011 (United Nations, 2011)

From the year of 1990 to 2010, the Chinese population increased from 1.14 billion to 1.341 billion. However, the population growth rate declined since 1990 according to the World Bank database (The World Bank, 2012c). Figure 5 shows the trend of Chinese population growth rate from 1980 to 2010; data taken from Appendix B, Table B. The total population increased slowly, but the population growth rate decreased year by year since 1990 due to the decrease in fertility and mortality (Cheng, Rosenberg, Wang, Yang, & Li, 2011). The reduction in population growth rate speeds up the growth in the aging population. An important need of the "aging society" is high quality healthcare, because the elderly are experiencing increasing rates of chronic diseases (Grimard, Laszlo, & Lim, 2010). Therefore, the medical device market in China is set to expand.

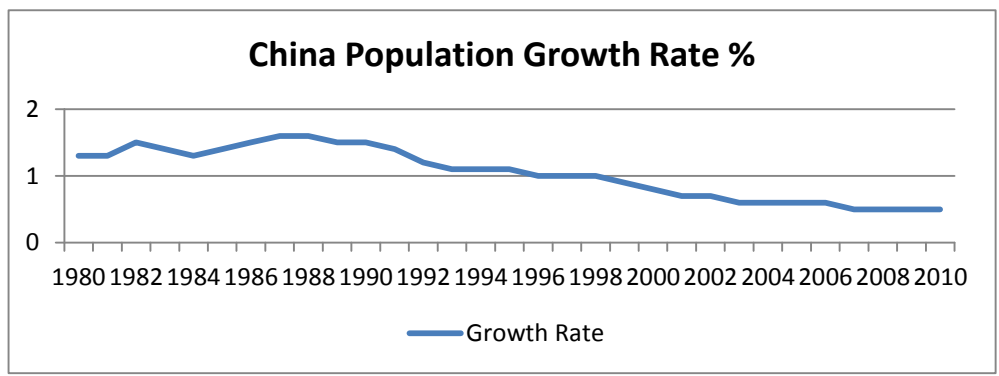

**Fig. 5.** China population growth rate from 1980 to 2010.
Source: The World Bank, 2012 (The World Bank, 2012c)

*4.2.2.  Neural networks time series prediction*

Prediction is used in a number of areas. Recently there has been a growing interest in applying neural networks to dynamic systems identification, prediction and control (Pham & Liu, 1995). Back Propagation (BP) Neural Networks are widely used with great success. McClelland, Rumelhart and Hinton (Rumelhart, McClelland, & University of California San Diego. PDP Research Group., 1986) established the Parallel Distributed Processing (PDP) models. The PDP research group proposed the BP algorithm, this method solved the problem of a lack of suitable training methods for the multilayer perceptron (MLP), and this greatly assisted the development of neural networks.

Neural networks are increasingly used by business and management for prediction and systems optimization. This includes financial analysis and forecasting (Sharda & Patil, 1992), bankruptcy prediction (Odom & Sharda, 1990) and stock market prediction (Mohan & Indian Institute of Management Ahmedabad., 2005). Luo and Huang (R. Luo & Huang;, 2004), Xie and Li (Xie & Y. Li, 2009) have discussed the prediction of population based on neural networks.

We use neural network time series prediction in this article. This is a simple way to predict population. In the neural network time series prediction, the method of nonlinear autoregression (NAR) (MathWorks MATLAB, 2011a) has been chosen because there is only one series involved (here it is Population). The future value schema of a time series *y(t)* is shown in Figure 6.

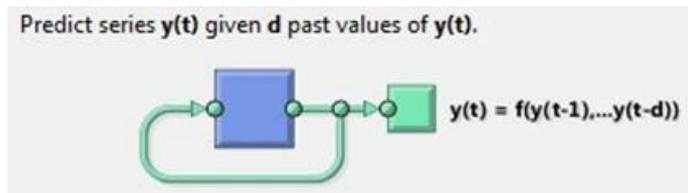

**Fig. 6.** Diagram of NAR neural network. (MathWorks MATLAB, 2011a)

Using a neural network time series prediction based on the real population from 1980 to 2010, the predicted total population in China from 2011 to 2020 is extrapolated in Figure 7. Data is shown in tabular form in Appendix C, Table C.1.

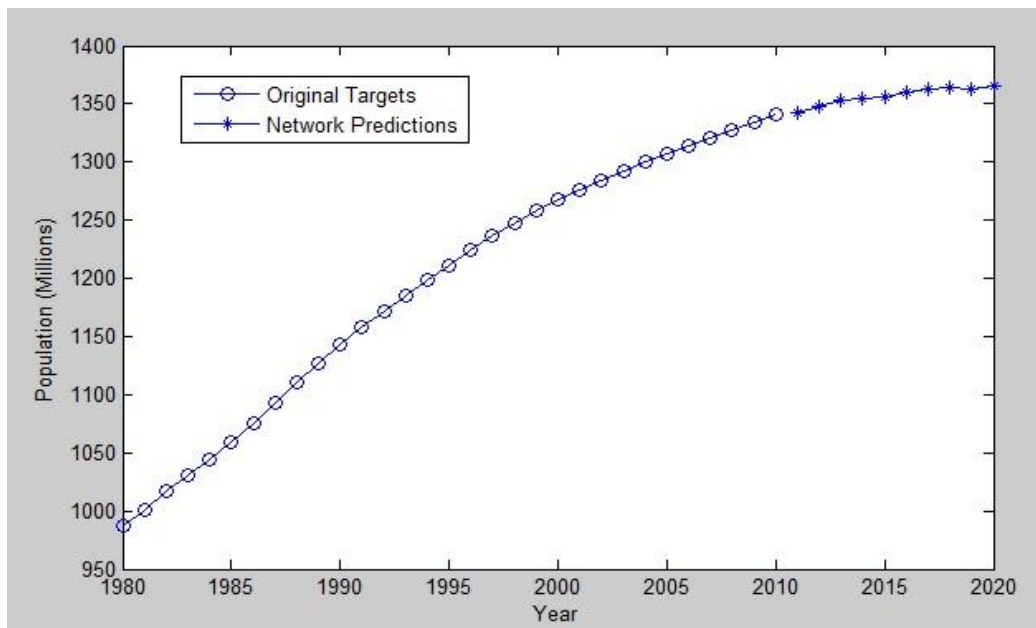

**Fig. 7.** Real Chinese population and predicted population.

After training many times, it was found that the optimum number of delays is 5, and the number of hidden neurons are 16 (see Figure 8), this minimized the error of the neural network[7] to 0.029528, which gives the best prediction performance. Detailed data is shown in Appendix C, Table C.2.

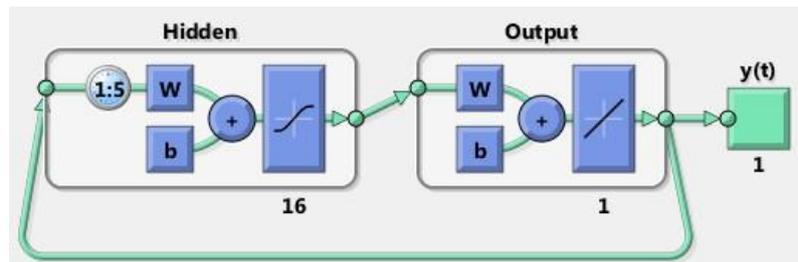

---

[7] Neural network error= $\sqrt{\sum_{i=1980}^{2010}(P_i - D_i)^2}$ ,where $P_i$ is year i population, $D_i$ is year i predicted population.

**Fig. 8.** Diagram of NAR neural network used for Chinese total population prediction.

The data shows that the Chinese total population in 2011 and 2012 is 1,344.13 million and 1,350.70 million respectively, while the results of the neural network time series prediction shows the population is 1,342.32 million (2011) and 1,347.63 million (2012) (The World Bank, 2013). The error is relatively small and a reliable prediction has been achieved. Therefore, a Neural Network performs well in predicting the population. This method is used to predict the 65 year old and above population in China, with results illustrated in Figure 9.

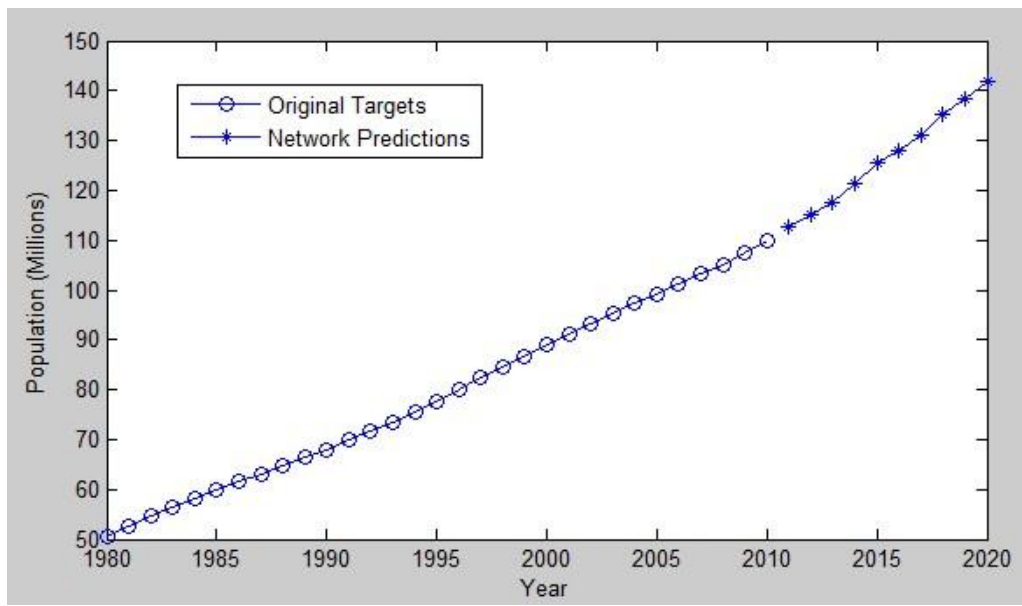

**Fig. 9.** Real Chinese 65+ population and predicted 65+ population

Predicting the population for 65 year olds and above permits prediction of a major element of the future value of the medical device market based on past records. Fig. 9 illustrates the real Chinese aged 65 and over population from 1980 to 2010 and the predicted 65 year olds and above population from 2011 to 2020. The predicted Chinese 65 year olds and above population from 2011 to 2020 is reported in Appendix C, Table C.1.

*4.3. China total healthcare expenditures and the number of hospitals*

According to the regression results, although China's total healthcare expenditures and the number of hospitals have a positive relationship with medical devices revenues; compared with the number of hospital visits and 65+ population, the contributions are relatively small, as they cause a positive change of 0.07 and 0.02 in medical devices revenues respectively.

China now has hundreds of thousands of medical and healthcare institutions; increasing numbers of hospital visits; unmet medical needs has made these medical and health institutions require more and more medical devices. China has a large population, who have experienced rapid economic development, which has resulted in continuous improvement of people's living standards; the descending population growth rate in China, means that the aging population has become a vital element in driving the medical device market. China's medical device industry is in a period of development and expansion, there are very broad market opportunities. The huge demand provides a good market opportunity for medical device companies. The number of medical and health institutions, and visits and inpatients in health institutions has experienced growth every year; consequently there is an increased demand for medical devices, especially for good quality, multi-functional medical systems. The frequency of use of medical devices will accelerate and the renewal period will be shortened. The improvement of Chinese medical and health services provides a good development platform and market capacity for the Chinese medical device industry.

We used quantitative methods to evaluate China's medical devices market investment potentials, but we cannot use the quantitative methods to calculate the other main important drivers of the Chinese medical device market, such as the ownership rate of medical devices among the main hospitals and disease profiles.

### 4.4. The ownership rate of medical devices among the main hospitals

The China Health Statistical Yearbook summarized the number of medical devices in the main hospitals and the percentage of medical devices in the main hospitals in China before 2004. Although no official data was collected after 2004, we can get some data from other agents such as CAME and CHA in order to explore the medical device market investment situation in China. Table 4 and Table 5 show the number and percentage of medical devices in the main hospitals in China.

**Table 4**
Number of medical devices in the main hospitals in China (units).

| Device Name / Year      | 1996   | 1998   | 2000   | 2001   | 2004   |
| ---                     | ---    | ---    | ---    | ---    | ---    |
| Electrocardiograph      | 35,295 | 41,230 | 46,122 | 48,073 | ---    |
| B-mode ultrasound       | 19,077 | 21,842 | 23,911 | 24,893 | 19,653 |
| Color Doppler ultrasound| 2,455  | 4,596  | 5,110  | 5,926  | 7,613  |
| CT                      | 2,549  | 3,543  | 4,247  | 4,760  | 4,752  |
| MRI                     | 356    | 512    | 604    | 714    | 1,110  |
| Cardiac Monitor         | 19,108 | 27,580 | 39,995 | 47,024 | ---    |

Source: China Health Statistics, (Ministry of Health of the People's Republic of China, 2004a, 2007a)

**Table 5**

Percentage of medical devices in the main hospitals in China (%).

| Device Name / Possession Rate | 1996 | 1998 | 2000 | 2001 | 2004 |
|---|---|---|---|---|---|
| Electrocardiograph | 88.1 | 90.2 | 92.8 | 93.7 | --- |
| B-mode ultrasound | 87.1 | 89.3 | 91.8 | 92.7 | 83.0 |
| Color Doppler ultrasound | 15.4 | 22.0 | 29.0 | 32.8 | 35.7 |
| CT | 17.8 | 22.6 | 27.7 | 30.6 | 29.2 |
| MRI | 2.5 | 3.2 | 4.0 | 4.8 | 7.2 |
| Cardiac Monitoring | 39.3 | 43.4 | 48.2 | 49.8 | --- |

Source: China Health Statistics, (Ministry of Health of the People's Republic of China, 2004a, 2007a)

Nearly 90% of the main hospitals purchased Electrocardiograph and B-mode ultrasound, because electrocardiograph and B-mode ultrasound are relatively cheap and are used widely, ordinary people can afford the diagnostic fees. Moreover, many of electrocardiograph and B-mode ultrasound devices are indigenous products, there is massive competition especially price competition in this area, which has reduced the profit margins, investment in these devices will not gain more profits. Table 4 illustrates that the units for Cardiac Monitoring were 47,024 in 2001, almost the same as Electrocardiograph, but its percentage in the main hospitals accounts for 49.8%, this is a large difference between Electrocardiograph's 93.7% in 2001. It means that with the high growth rate of the number of Cardiac Monitors, according to Table 4, the market for Cardiac Monitors still has room for expansion. From both Table 4 and Table 5 we can understand that the market for Color Doppler ultrasound, CT and MRI has great investment potential, especially MRI. This situation is caused for many reasons including the disease profiles, which are described below.

Today the hospital's reputation is said to rely on them possessing the latest and most expensive medical devices such as Color Doppler ultrasound, CT and MRI (Schulze, 2001), because most hospital's revenues are generated by these expensive medical devices. In 1996, China had 2,549 CT scanners, 2,455 Color Doppler ultrasound and 356 MRIs. The increasing speed of adoption of these medical devices is impressive. Until 2004, China had 4,752 CT scanners, 7,613 Color Doppler ultrasound and 1,110 MRIs (see Table 4). According to CHA's report (Chinese Hospital Association, 2012), China had 9,109 CT scanners in 2008, 10,101 CT scanners in 2009 and 11,242 CT scanners in 2010. With the increasing number of CT's in China, China had 5.5 CT scanners per million people in 2006, which increased to 8.6 CT scanners per million people in 2010, which shows a rapid growth. However, compared with other countries, the ownership rate of CT devices is relatively low. For example, Japan had 98 CT scanners per million people in 2006. Therefore, it can be predicted that there will be huge future demand for this kind of medical device in China.

*4.5. The main diseases in China*

In China, with the increase in the population of elderly people; the improvement of people's living standards; population movements and the accelerated process of urbanization, disease profiles have altered significantly.

China now belongs to the upper middle income countries (The World Bank, 2012a); of the top ten leading causes of death in the middle income countries, seven are chronic disease-related deaths, which accounted for 91% of total deaths (World Health Organization, 2008). The higher burden of chronic diseases in low- and middle income countries is well described by China (Merson, Black, & Mills, 2012), which means these diseases will cost a great deal. Although digestive diseases, respiratory diseases, infectious and parasitic diseases are the top ten leading cause of death in low- and middle income countries, we need to focus more attention on cancers, cardiovascular diseases and cerebrovascular diseases, which account for the top three percent of total deaths in China (World Health Organization, 2008).

Figure 10 and Figure 11 show the percentage of total deaths from the top five main diseases in Chinese cities and counties from 2003 to 2011. All the data are collected from the Ministry of Health of the People's Republic of China health statistics. Detailed data is shown in Appendix A, Table A.1. (city) and Appendix A, Table A.2 (county). Due to the scarcity of data; the years of 2007 and 2010 are not included.

Figure 10 illustrates the percentage of total deaths from the top 5 main diseases in China's cities from 2003 to 2011. By comparison, the top three leading causes of death in middle-income countries are cardiovascular diseases, cerebrovascular diseases and respiratory diseases (Beaglehole & Yach, 2003; Kirton, 2009; Merson, et al., 2012), there were 2.8 million deaths from cardiovascular diseases in China in 2003 (Beaglehole & Yach, 2003). China has a different profile when compared with other middle-income countries. Cancers caused the highest mortality in China and has maintained the first position among the five leading causes of death in both the cities and counties of China, except in 2005, according to Figure 11. The major risk factors causing cancers are tobacco consumption, chronic infections, diet and lack of physical activity, etc. (Stewart, Kleihues, & International Agency for Research on Cancer., 2003).

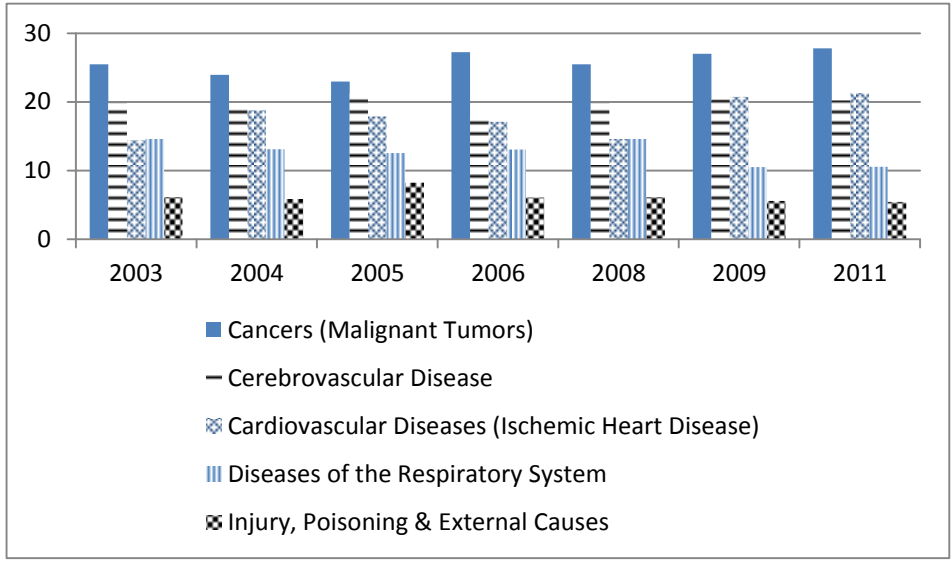

**Fig. 10.** Percentage of total deaths from the top 5 main diseases in Cities
Source: China Health Statistics, (Ministry of Health of the People's Republic of China, 2004b, 2005, 2006, 2007b, 2008, 2009, 2010, 2012)

Compared with Figure 10 and Figure 11, cancers, cardiovascular diseases and cerebrovascular diseases are the top three causes of death in cities in China, but we cannot neglect respiratory diseases, which accounted for 23.45% of deaths in the counties of China in 2005 (see Appendix A, Table A.2.). Major risk factors for respiratory diseases include air pollution, tobacco consumption, occupational long term exposures, etc. (Ait-Khaled, Enarson, & Bousquet, 2001). Cerebrovascular diseases have reached second place among the ten leading causes of death in the developing countries, as well as in China. Major risk factors for cerebrovascular diseases are tobacco consumption, obesity and life stress, etc. Injury, poisoning and external causes are ranked in position five of the leading causes of death in both the cities and counties in China. From analysis of these diseases, it can be seen that tobacco is the greatest cause of health problems. Several diseases and conditions were added to the lists as being causally related to smoking (U.S. Department of Health and Human Services (DHHS), 2004).

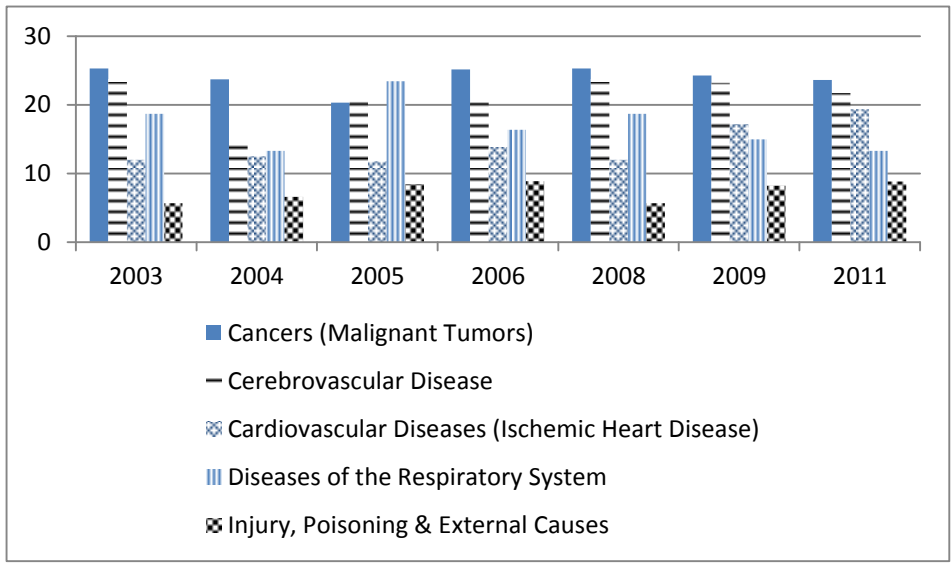

**Fig. 11.** Percentage of total deaths from the top 5 main diseases in the County regions. Source: China Health Statistics, (Ministry of Health of the People's Republic of China, 2004b, 2005, 2006, 2007b, 2008, 2009, 2010, 2012)

Cancer is a leading cause of death globally, accounting for 7.6 million deaths in 2008 (World Health Organization, 2013). Nearly 70% of cancer deaths occurred in low- and middle-income countries. It is predicted that deaths from cancer will increase, with an estimated 13.1 million deaths in 2030 (International Agency for Research on Cancer, 2010). Cancer is a big problem for the society worldwide as well as for China.

## 5. Discussion

Our data suggests that the Chinese medical device market is not only driven by the four variables (number of hospital visits; 65+ population; total healthcare expenditures and number of hospitals) that were discussed earlier but is also impacted by the ownership rate of medical devices and the diseases profile, for example cancer. Results from regression analysis suggest that aging (mainly for people aged 65 and above) is one of the important drivers of the Chinese medical device market. The correlation coefficient informs us about the direction and the relationship strength between the dependent variable and independent variables. Interestingly, total healthcare expenditure and hospital quantity are not the major factors (compared with 65+ population and number of hospital visits) in controlling the Chinese medical device market but have a positive correlation. Furthermore, our data indicates that the ownership rate of medical devices among the main hospitals and disease profiles are important (but not easily quantified) in driving the medical device market.

Regression analysis illustrates that people aged 65 and above play an important role in China's medical device market. According to Gu's research, Chinese 65+ years population are projected to be 236 and 334 million by the years 2030 and 2050, respectively (Gu, Dupre, Warner, & Zeng, 2009). Therefore, due to the importance of the aging population in China, we used neural network time series prediction to estimate the total population and 65+ population from 2010 to 2020. Prediction results show that there is a smooth rising trend of Chinese 65+ population from 2010 to 2020, revealing that aging population plays an important role in the Chinese healthcare industry, which indicates that the Chinese medical device market has possible investment opportunities in the future.

The purpose of medical devices is to assist with: patient' stratification, diagnosis, prognosis, treatment and treatment planning; the macroeconomic variables such as population structure; disease profiles and economic level can affect the demand for medical device services. Disease profiles affect the development of medicine as well as medical device capabilities and the total medical device market. Therefore, diseases should be one of the elements driving medical device industrial investment. The top five diseases have a significant impact on China's economy via healthcare costs and lost productivity. Popkin et al (Popkin, Horton, Kim, Mahal, & Shuigao, 2001) estimated that diet-related chronic diseases (cancer, cardiovascular diseases, cerebrovascular diseases and diabetes, etc.) accounted for 22.6% of healthcare costs in China, while the cost of lost productivity due to these diseases was about 0.5% of GDP in 1995. Cancer is still a big challenge for the world as well as for China. The top five cancers (Lung, liver, stomach, oesophagus and colorectal) in China in 2004 still maintained their position in 2008; all cancers have a smooth rising trend except breast cancer, which shows an rapid increase of 4.02% in mortality rate from 2004 (2.88%) to 2008 (6.9%) (Ministry of Health of the People's Republic of China, 2012). Among the main cancers in China, breast cancer is a growing problem; further investment in research and facilities to combat breast cancer in China is recommended.

If the incidence or mortality from the disease is low, the demand and frequency of use of the appropriate diagnosis and treatment equipment will be relatively low, the investment payback period for such medical devices will be long for hospitals; in such a scenario, it is difficult for hospitals to recover the cost of medical devices throughout their entire life cycle. So only the large general hospitals will consider purchasing such medical devices. Small and medium-sized hospitals do not have the capacity to buy such medical devices. Thus, the market demand for medical devices with low disease incidence is relatively small; investment risk is large and does not have investment value. If the incidence or mortality of the disease is high, the demand and frequency of use of the appropriate diagnosis and treatment equipment will be relatively high, the large general hospitals will be very motivated to purchase such medical devices as well as small and medium-sized hospitals because the investment payback period for such devices will be short.

With the advancement of medical technology, the diseases which threaten human life are changing as well. Cancers, cerebrovascular diseases, cardiovascular diseases and respiratory diseases will become the major diseases which threaten human health and life in the 21$^{st}$ century. With the gradual increase in the number of patients and mortality, the demand for diagnosis and treatment devices will inevitably increase. Good market prospects indicate that medical devices for these diseases have great investment potential.

Medical devices demand analysis is the basis of industrial investment opportunities. This study predicts the medical device market demand based on disease profiles and the extent of the medical device market saturation. In a perfectly competitive market, the prices of all products and services in an industry are determined by equating the demand for a good with its supply (Cohen & Winn, 2007). In the economics area, there is a principle that as "the price of a good or service rise, the quantity demand falls", this is the *price elasticity of demand*. However, research reports that the demand for medical technology is extremely inelastic (it does not rely on price) and free of this principle "competition will make prices fall, narrowing margins and reducing profits" (Kruger, 2005). Other arguments show that the demand for most new products tends to be based on non-price factors (Hill, 2009). The medical device market has sustainable growth because of the general demographic trends, especially the growth of the aging population and the continued prevalence of diseases (Kruger, 2005). For the Chinese medical device market, the growth of medical services institutions, medical diagnosis and treatment devices, disease profiles and population - especially the aging population, means that the market has great investment potential.

The main medical devices companies' investment activities and/or mergers and acquisitions in China give good indicators of how the market is developing. For example: Philips Healthcare (Philips Healthcare, 2012) established medical imaging bases in Suzhou (Jiangsu Province) in 2009 and purchased Shanghai Apex Electronics in 2010 for ultrasound production. GE Healthcare (General Electric, 2012) is to form a joint venture with Shinva Medical Instrument in Zibo (Shandong Province), which was the first medical devices company in China (established in 1943). GE Medical Industrial Park (Beijing) is one of the largest research and development (R&D) bases in the world. The partnership, aims to be active in R&D, and manufacturing/sales, the production of medical devices includes: Computed Tomography (CT), Magnetic Resonance Imaging (MRI) and X-rays. The revenues from GE Healthcare was one billion US dollars in China in 2008 with increasing sales performance. Siemens Healthcare (Siemens Healthcare, 2012) has two wholly owned subsidiaries in China, one named Siemens Shanghai Medical Equipment Ltd. (founded in 1992) in Shanghai, which is a research and development base and manufactures CT and X-ray systems; the other one is Siemens Shenzhen Magnetic Resonance Ltd. (founded in 1998) in Shenzhen (Guangdong Province). Hitachi (Hitachi Medical Systems, 2012) established three Medical Systems Corporations in Suzhou, Beijing and Guangzhou. Toshiba Medical System (China) Corporation Ltd was established in Beijing in 2007 for developing CT devices. All these activities contribute to China's medical device industrial growth. Moreover, the Chinese government has developed rural areas healthcare systems in recent years and this investment is continuing, this creates more investment potential for investors. Therefore, foreign investment can make a positive contribution to a host country by supplying capital, management resources and technology that would otherwise not be available and this increases the country's economic growth rate (Li & Liu, 2005; Lipsey, 2002); enhanced technology prowess can stimulate further economic development and industrialization (Hill, 2009). So, China is a market with fabulous investment potential for health care providers and medical device manufacturers (Schulze, 2001).

  As with many studies, this study has several limitations. Some of the data from the government report is hard to access, for example: the number and percentage of medical devices in the main hospitals in China, for which data was only reported for the years of 1996, 1998, 2000, 2001 and 2004. Trying to use questionnaires is unreliable. Generally speaking, the scarcity of data does not have a big impact on the main results. Data deficiency is a major problem faced by most countries. The number of medical devices in Chinese hospitals after 2004 is hard to access from the Chinese Ministry of Health database. We selected four important drivers of the Chinese medical device market, because we can access this data to use quantitative methods to assess the market. The market's drivers not only include the four drivers and the ownership rate of medical devices and disease, but also contains other elements. Despite the data limitations, the article describes China's medical devices current market situation and identifies the investment potential of the market. Analysis of the prevalence of diseases in China shows that cancers are the big challenge for the whole medical area, and that there is an increasing trend in the incidence of death from breast cancer, this indicates that there is a requirement for further research and analysis in this area.

## 6. Conclusions

This study indicates that the Chinese medical device market has great investment potential for identified reasons. China has a rapidly developing economy, there is government investment in the healthcare industry to improve the medical environment, the government is encouraging foreign medical companies to investment in China. Some large foreign companies like GE healthcare, Siemens healthcare and Philips healthcare continue to increase investment in China. This indicates that the Chinese medical device market has significant investment potential for the future. Foreign investment will bring both benefits and risks to the country's health sector (Smith, 2004). The Chinese medical device market still has room for investment due to the growing aging population and the increasing number of hospital visits. Because China continues to develop economically and socially, the predominance of infectious diseases is decreasing. Chronic diseases (epidemiological transition) are emerging as an increasing problem, hence the health care system is experiencing huge pressures from both changing and increasing demands (Dummer & Cook, 2008).

This study suggests that the Chinese medical device market has great potential and shows that medical diagnosis and treatment devices will be in tremendous demand in the future. The rising numbers of aging people in China, the changed disease profiles and the constant increase in the incidence of chronic diseases like cancers, which requires medical diagnosis and treatment devices such as CT, MRI and ultrasound are driving this demand.

It is noteworthy that China's high-end medical device market relies on imports from the developed countries. These foreign made high-tech medical devices account for 70% of China's medical device market. The number of medical and health institutions increased year by year as did the number of visits and in-patients in healthcare institutions, this huge and increasing demand provides an expanding market for medical devices. Disease profiles determine which kind of medical devices have more investment value in China. Demand analysis indicates that Color Doppler ultrasound, CT and MRI have great future investment potential in China. The growth in the aging population is a big test of the health care industry in China, peoples' desire for good health is stronger than before, therefore significant opportunities exist in the medical device market.

This article provides the first study of its kind in providing a better understanding of China's medical device market. We hope our study has shed some light on this important topic and we encourage more studies for this important market opportunity.

**Appendix A.**

*Table A.1. Percentage of Total Deaths of Top 5 Main Diseases in Certain Region from 2003 to 2011(City)*

| Cause / % | 2003 | 2004 | 2005 | 2006 | 2008 | 2009 | 2011 |
|---|---|---|---|---|---|---|---|
| Cancers (Malignant Tumors) | 25.47 | 23.92 | 22.94 | 27.25 | 25.47 | 27.01 | 27.79 |
| Cerebrovascular Disease | 19.95 | 19.09 | 21.23 | 17.66 | 19.95 | 20.36 | 20.22 |
| Cardiovascular Diseases (Ischemic Heart Disease) | 14.43 | 18.80 | 17.89 | 17.10 | 14.63 | 20.77 | 21.30 |
| Diseases of the Respiratory System | 14.63 | 13.12 | 12.57 | 13.06 | 14.63 | 10.54 | 10.56 |
| Injury, Poisoning & External Causes | 6.16 | 5.89 | 8.25 | 6.10 | 6.16 | 5.59 | 5.47 |

Source: China Health Statistics, (Ministry of Health of the People's Republic of China, 2004b, 2008, 2010, 2012)

*Table A.2. Percentage of Total Deaths of Top 5 Main Diseases in Certain Region from 2003 to 2011(County)*

| Cause / % | 2003 | 2004 | 2005 | 2006 | 2008 | 2009 | 2011 |
|---|---|---|---|---|---|---|---|
| Cancers (Malignant Tumors) | 25.28 | 23.70 | 20.29 | 25.14 | 25.28 | 24.26 | 23.62 |
| Cerebrovascular Disease | 23.75 | 14.85 | 21.17 | 20.36 | 23.75 | 23.19 | 21.72 |
| Cardiovascular Diseases (Ischemic Heart Disease) | 12.03 | 12.54 | 11.77 | 13.87 | 12.03 | 17.21 | 19.37 |
| Diseases of the Respiratory System | 18.72 | 13.30 | 23.45 | 16.40 | 18.72 | 14.96 | 13.31 |
| Injury, Poisoning & External Causes | 5.69 | 6.63 | 8.47 | 8.90 | 5.69 | 8.25 | 8.85 |

Source: China Health Statistics, (Ministry of Health of the People's Republic of China, 2004b, 2008, 2010, 2012)

**Appendix B.**

*Table B. Chinese population and its relevant data from 1980 to 2010*

| Year/Ages | China total 65 and above population (millions) | China total population (millions) | Percentages of 65 and above population in China (%) | China total population growth rate (annual %) |
| --- | --- | --- | --- | --- |
| 1980 | 50.677 | 987.05 | 5.13 | 1.3 |
| 1981 | 52.697 | 1,000.72 | 5.27 | 1.3 |
| 1982 | 54.594 | 1,016.54 | 5.37 | 1.5 |
| 1983 | 56.419 | 1,030.08 | 5.48 | 1.4 |
| 1984 | 58.218 | 1,043.57 | 5.58 | 1.3 |
| 1985 | 60.009 | 1,058.51 | 5.67 | 1.4 |
| 1986 | 61.565 | 1,075.07 | 5.73 | 1.5 |
| 1987 | 63.149 | 1,093.00 | 5.78 | 1.6 |
| 1988 | 64.754 | 1,110.26 | 5.83 | 1.6 |
| 1989 | 66.377 | 1,127.04 | 5.89 | 1.5 |
| 1990 | 68.05 | 1,143.33 | 5.95 | 1.5 |
| 1991 | 69.808 | 1,158.23 | 6.03 | 1.4 |
| 1992 | 71.671 | 1,171.71 | 6.12 | 1.2 |
| 1993 | 73.608 | 1,185.17 | 6.21 | 1.1 |
| 1994 | 75.58 | 1,198.50 | 6.31 | 1.1 |
| 1995 | 77.576 | 1,211.21 | 6.40 | 1.1 |
| 1996 | 80.073 | 1,223.89 | 6.54 | 1 |
| 1997 | 82.387 | 1,236.26 | 6.66 | 1 |
| 1998 | 84.584 | 1,247.61 | 6.78 | 1 |
| 1999 | 86.749 | 1,257.86 | 6.90 | 0.9 |
| 2000 | 88.912 | 1,267.43 | 7.02 | 0.8 |
| 2001 | 91.044 | 1,276.27 | 7.13 | 0.7 |
| 2002 | 93.202 | 1,284.53 | 7.26 | 0.7 |
| 2003 | 95.336 | 1,292.27 | 7.38 | 0.6 |
| 2004 | 97.312 | 1,299.88 | 7.49 | 0.6 |
| 2005 | 99.087 | 1,307.56 | 7.58 | 0.6 |
| 2006 | 101.237 | 1,314.48 | 7.70 | 0.6 |
| 2007 | 103.21 | 1,321.29 | 7.81 | 0.5 |
| 2008 | 105.163 | 1,328.02 | 7.92 | 0.5 |
| 2009 | 107.325 | 1,334.50 | 8.04 | 0.5 |
| 2010 | 109.845 | 1,340.91 | 8.19 | 0.5 |

**Appendix C.**

*Table C.1. The predicted Chinese population from 2011 to 2020*

| Year/Ages | Total China population (millions) | Total 65+ population (millions) |
|---|---|---|
| 2011 | 1,342.32 | 112.71 |
| 2012 | 1,347.63 | 115.14 |
| 2013 | 1,353.85 | 117.61 |
| 2014 | 1,354.77 | 121.50 |
| 2015 | 1,356.35 | 125.45 |
| 2016 | 1,359.52 | 127.99 |
| 2017 | 1,362.93 | 131.16 |
| 2018 | 1,364.14 | 135.08 |
| 2019 | 1,363.05 | 138.37 |
| 2020 | 1,365.06 | 141.98 |

*Table C.2. Neural networks*

| Neurons | Neural Network Error (res) |
|---|---|
| 4 | 0.049233 |
| 5 | 0.16097 |
| 6 | 0.048822 |
| 7 | 0.1729 |
| 8 | 0.078508 |
| 9 | 0.114535 |
| 10 | 0.171799 |
| 11 | 0.105598 |
| 12 | 0.166333 |
| 13 | 0.127967 |
| 14 | 0.091731 |
| 15 | 0.047639 |
| 16 | 0.029528 |
| 17 | 0.070366 |
| 18 | 0.111514 |